\pdfoutput=1
\documentclass[%
superscriptaddress,
preprint,
 amsmath,amssymb,
 aps,
prb,
]{revtex4-2}

\usepackage{graphicx}% Include figure files
\usepackage{dcolumn}% Align table columns on decimal point
\usepackage{bm}% bold math
\usepackage{multirow}
\usepackage{siunitx}
\usepackage{amsmath}
\usepackage{amssymb}
\usepackage{bbold}
\usepackage{color}
\usepackage{comment}
\usepackage{tikz}
\usepackage{pgfplots}
\pgfplotsset{compat=1.5}
\usepackage{graphicx}
\usepackage{blindtext}
\usepackage{physics}
\usepackage{xcolor}
\usepackage{ulem}
\usepackage{soul}  % For highlighting text
\usepackage[right,mathlines]{lineno}

\definecolor{dodgerblue}{HTML}{1E90FF}
\newcommand{\moire}{moir\'e}
\newcommand{\Moire}{Moir\'e}

\usepackage[colorlinks=true,linkcolor=black,citecolor=dodgerblue, urlcolor=dodgerblue]{hyperref}

\renewcommand{\figurename}{\textbf{Fig.}}

\newcommand{\Rmnum}[1]{\expandafter\@slowromancap\romannumeral #1@}

\begin{document}

%\linenumbers 

\title{Mapping the {\moire} potential in multi-layer rhombohedral graphene}

\author{Eric Seewald}
\thanks{These authors contributed equally}
\affiliation{Department of Physics, Columbia University, New York, NY, USA}
\author{Sanat Ghosh}
\thanks{These authors contributed equally}
\affiliation{Department of Physics, Columbia University, New York, NY, USA}
\author{Nishchhal Verma}
\thanks{These authors contributed equally}
\affiliation{Department of Physics, Columbia University, New York, NY, USA}
\author{John Cenker}
\thanks{These authors contributed equally}
\affiliation{Department of Physics, Columbia University, New York, NY, USA}
\author{Yinan Dong}
\affiliation{Department of Physics, Columbia University, New York, NY, USA}
\author{Birui Yang}
\affiliation{Department of Physics, Columbia University, New York, NY, USA}
\author{Amit Basu}
\affiliation{Department of Condensed Matter Physics and Materials Science, Tata Institute of Fundamental Research, Mumbai, India}
\author{Takashi Taniguchi}
\affiliation{National Institute for Materials Science, Tsukuba, Japan}
\author{Kenji Watanabe}
\affiliation{National Institute for Materials Science, Tsukuba, Japan}
\author{Mandar M. Deshmukh}
\affiliation{Department of Condensed Matter Physics and Materials Science, Tata Institute of Fundamental Research, Mumbai, India}
\author{Dmitri N. Basov}
\affiliation{Department of Physics, Columbia University, New York, NY, USA}
\author{Raquel Queiroz}
\affiliation{Department of Physics, Columbia University, New York, NY, USA}
\affiliation{Center for Computational Quantum Physics, Flatiron Institute, New York, NY, USA}
\author{Cory Dean}
\affiliation{Department of Physics, Columbia University, New York, NY, USA}
\author{Abhay N. Pasupathy} 
\email{apn2108@columbia.edu}
\affiliation{Department of Physics, Columbia University, New York, NY, USA}
\affiliation{Condensed Matter Physics and Materials Science Department, Brookhaven National Laboratory, Upton, NY, USA}

\date{\today}

\begin{abstract}
Rhombohedral graphene (rG) aligned with hexagonal boron nitride (hBN) has been shown to host flat bands that stabilize various strongly correlated quantum phases, including Mott insulators, integer, and fractional quantum anomalous Hall phases. 
In this work, we use scanning tunneling microscopy/spectroscopy (STM/STS) to visualize the dispersion of flat bands with doping and applied displacement fields in a hBN aligned rhombohedral trilayer graphene (rtG)/hBN {\moire} superlattice. 
In addition to the intrinsic flat bands of rtG induced by the displacement field, we observe low-energy features originating from {\moire} potential-induced band folding.
Real-space variations of the spectroscopic features allow us to quantify the spatial structure of the {\moire} potential at the rtG/hBN interface. Importantly, we find that accurately capturing the {\moire} site–dependent spectra requires incorporating a {\moire} potential acting on the top graphene layer with a sign opposite to that of the bottom layer into the continuum model. Our results thus provide key experimental and theoretical insights in understanding the role of the {\moire} superlattice in rG/hBN heterostructures.
\end{abstract}

\maketitle

\newpage

\begin{figure}
\includegraphics[width=16.8cm]{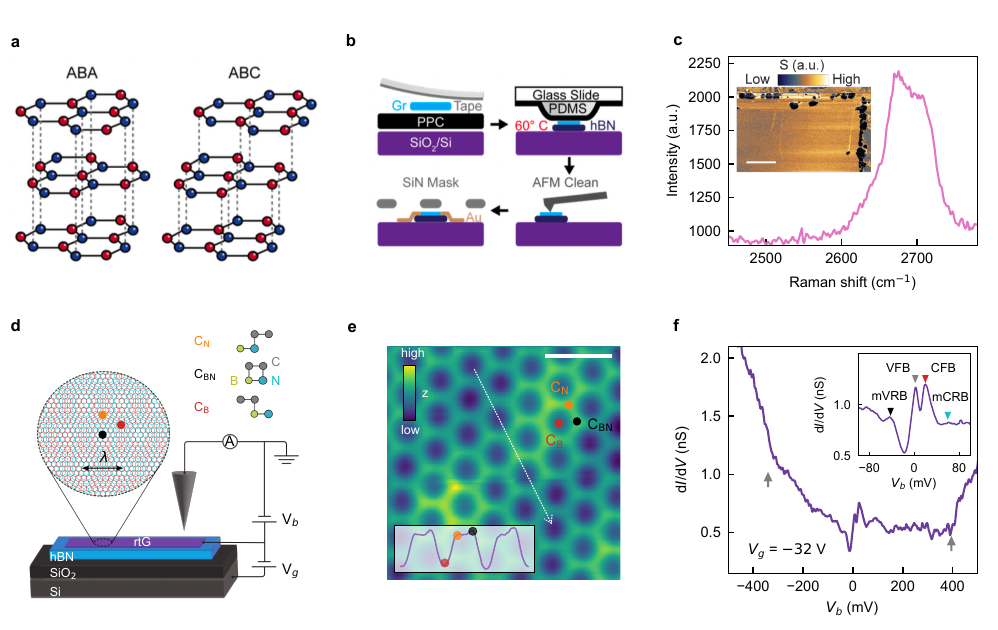}
\caption{\label{fig:fig1}\textbf{Stabilization and identification of large area trilayer rhombohedral (ABC) graphene.} (a) Bernal (ABA) and rhombohedral (ABC) stacking configurations in trilayer graphene. (b) Schematic showing fabrication steps of trilayer ABC graphene aligned with hBN. (c) Raman signal of the fabricated device at room temperature. Inset shows a large area scanning near field optical microscopy (SNOM) map showing uniformity of ABC phase. Scale bar 2 \textmu m. (d) Measurement schematic of scanning tunneling microscopy/spectroscopy (STM/STS) on trilayer ABC graphene aligned with hBN. Inset: the local stacking configuration within the hBN/graphene {\moire} of wavelength $\lambda$. Different sites C$_\mathrm{BN}$, C$_\mathrm{B}$, and C$_\mathrm{N}$ of the {\moire} are labeled. (e) Constant tunneling current (600 mV, 70 pA) topography of the hBN/ABC graphene {\moire} pattern. Scale bar $30$ nm. Inset: Line cut profile along the arrow showing the inequivalent C$_\mathrm{BN}$, C$_\mathrm{B}$, and C$_\mathrm{N}$ sites. (f) Measured d$I$/d$V$ spectra at site C$_\mathrm{B}$ (marked by red dot in (e)) at $7$~K and at charge neutrality ($V_\text{g}$ = $-32$ V) point. The positions of remote bands are marked by vertical arrows. Inset shows zoomed-in low energy spectra with higher energy resolution showing valence flat band (VFB), conduction flat band (CFB), {\moire} valence remote band (mVRB), and {\moire} conduction remote band (mCRB).}
\end{figure}

The electronic structure of multi-layer graphene is highly sensitive to the local stacking order \cite{Haering1958, min_electronic_2008, Koshino2013, zhang_experimental_2011}.
For three layers, the two primary stacking configurations are the thermodynamically stable Bernal (ABA) (Fig.~\ref{fig:fig1}a, left) and the metastable rhombohedral (ABC) (Fig.~\ref{fig:fig1}a, right), distinguished by the relative position of the third layer (A or C) with respect to the first two (AB). Rhombohedrally stacked graphene heterostructures exhibit flat bands that are tunable by an external displacement field \cite{xu_direct_2015, Pierucci2015, Henck2018} and host a variety of quantum phenomena such as superconductivity \cite{zhou_superconductivity_2021, Han2024, Yang2024}, half- and quarter-metallic states \cite{zhou_half-_2021}, correlated insulating phases \cite{han_correlated_2024, liu_spontaneous_2024}, and valley ferroelectricity \cite{han_orbital_2023}. 
In addition to the rich correlated physics intrinsic to bare rhombohedral structure, introducing a periodic {\moire} potential through alignment to an hBN substrate generates a variety of gapped states at specific filling fractions in rtG \cite{Chittari2019, Gonzalez2021a}. Prominent among these are the recent observation of integer and fractional quantum anomalous Hall states in rhombohedral pentalayer graphene \cite{lu_fractional_2024}.
However, there are many open questions, such as whether the {\moire} potential is confined to the bottom layer or affects the top layer as well, whether the {\moire}'s influence on the top layer originates from single-particle effects or electron interactions, and how the {\moire} potential contributes to stabilizing correlated phases \cite{Zhou2024moire, Dong2024moire, Soejima2024moire, Kwan2025, Parameswaran2024AnomalousHallCrystals}.

To address these issues, we investigate the gate-dependent electronic properties of a rtG/hBN {\moire} heterostructure in real space using a combination of STM/STS and continuum model calculations. Such a measurement is made possible through a novel fabrication method that preserves the rhombohedral stacking order of few-layer graphene during lamination with hBN. By combining spectroscopy at different {\moire} stacking sites with continuum model calculations, we identify the hBN alignment orientation responsible for generating the specific {\moire} potential profile at the rtG/hBN interface in our device. We further demonstrate that reproducing the spectroscopic features using single-particle theory at different {\moire} sites requires incorporating a {\moire} potential of at least $- 30 \%$ of the bottom layer moire strength onto the top graphene layer.

Few-layer rhombohedral graphene exhibits exceptional transport properties, yet its real-space electronic structure has remained largely unexplored by scanning tunneling microscopy and spectroscopy (STM/STS). Previous studies have been limited to nitrogen temperatures \cite{Zhang_NN20_2025}, and those at helium temperatures have lacked the superlattice potential from hBN \cite{liu2024_IVC}. One major challenge is posed by the metastable nature of the rhombohedral stacking. Strain is believed to play a dominant role in the instability of rtG, as it has been shown to be thermally stable up to $800^\circ$~C \cite{lui_imaging_2011} but nevertheless tends to revert fully to Bernal stacking during sample fabrication. While van der Waals heterostructures are commonly assembled by sequentially picking up individual flakes using an adhesive polymer such as polypropylene carbonate (PPC), the mechanical stress induced in the flake during pick-up is suspected to convert the rhombohedral stacking to Bernal unless the domains are first isolated via anodic oxidation or reactive ion etching \cite{zhou_half-_2021}. This tendency, combined with reports that virtually all exfoliated flakes have Bernal domains \cite{zhang2020light}, poses significant challenges in the fabrication of rtG/hBN samples, especially open-faced devices with atomically clean surfaces required for STM measurements. Here we report a method to overcome this fabrication challenge.

The fabrication procedure of our device is shown schematically in 
Fig.~\ref{fig:fig1}b. First, bulk graphite crystals are exfoliated onto PPC spin-coated standard Si/${\rm SiO}_2$ substrates. Atomically thin graphite flakes with large rhombohedral domains are identified on the PPC film by their optical contrast and Raman spectra. The PPC film is then peeled off of the silicon substrate and placed on a poly(dimethylsiloxane) (PDMS) cylinder. The graphene is aligned with a bottom hBN layer and released from the PPC at $60^\circ$~C. We hypothesize that dropping the flake onto hBN (instead of picking up from silicon substrate with hBN) at such low temperatures minimizes anisotropic mechanical and thermal strain variations within the flake during stacking which tend to revert ABC domains to ABA \cite{guerrero2022rhombohedral}.

The surface of the heterostructure is then cleaned via gentle contact-mode AFM scans. The device is completed by directly depositing metal contacts (Cr/Au) through a SiN shadow mask \cite{deshmukh_nanofabrication_1999}. Fig.~\ref{fig:fig1}c shows the Raman spectra and scanning near field optical microscopy (SNOM) map (inset) of a completed, open-faced rtG/hBN device at room temperature. The characteristic asymmetric peak shape of the 2D-mode in the Raman spectra \cite{lui_imaging_2011}, combined with the uniformity of the SNOM map, suggests that the large, cleaned area of the sample maintains rhombohedral stacking throughout the fabrication process. 

After initial Raman and SNOM characterization at room temperature, we examined the low-energy electronic structure using differential tunneling conductance (d$I$/d$V$) spectroscopy at low temperature, as shown in the measurement schematic in Fig.~\ref{fig:fig1}d. For d$I$/d$V$ measurements, a \textit{dc} bias voltage ($V_\text{b}$) was applied to the graphene sample relative to the STM tip, together with a small \textit{ac} modulation voltage. The underlying Si substrate and graphene device form a parallel plate capacitor with SiO$_2$ and hBN serving as the dielectric. In this configuration, applying a back gate voltage ($V_\text{g}$) to the Si substrate simultaneously modulates the carrier density ($n$) and the vertical displacement field ($D$) in the rtG. We note that the lack of independent control over $D$ and $n$ in STM/STS measurements restricts the accessible parameter space compared to transport studies with a dual gate setup. Instead, in our measurement geometry, the STM tip acts as a local gate that affects the local doping and displacement field in rtG, as discussed later. All STS/STM measurements are performed at the base temperature $T \approx 7$ K of the system.

Fig.~\ref{fig:fig1}e presents a constant-current (600 mV, 70 pA) topography of the rtG/hBN heterostructure, revealing a {\moire} pattern with a periodicity of $\approx 14$~nm, consistent with near-perfect crystallographic alignment between rtG and hBN. The pattern remains uniform over hundreds of nanometers, indicating high surface quality and minimal polymer residues. Within each {\moire} unit cell, three distinct stacking configurations: C$_\mathrm{BN}$, C$_\mathrm{B}$, and C$_\mathrm{N}$, are identified from the relative height variations along the profile shown in the inset of Fig.~\ref{fig:fig1}e. These variations arise from out-of-plane deformations driven by local repulsive interactions between carbon atoms in the bottom graphene layer and boron/nitrogen atoms in the top hBN layer \cite{Zhou2015_vdwEnergetics, Elder_Huang_AlaNissila_2023}. At the C$_\mathrm{BN}$ site, carbon atoms lie directly above both boron and nitrogen atoms, whereas at C$_\mathrm{B}$ (C$_\mathrm{N}$) sites they lie above only boron (nitrogen) atoms, as illustrated schematically in the inset of Fig.~\ref{fig:fig1}d. The stacking energy is highest for C$_\mathrm{BN}$ and lowest for C$_\mathrm{B}$, with C$_\mathrm{N}$ slightly lower than C$_\mathrm{BN}$ \cite{argentero_unraveling_2017, moore_nanoscale_2021}. We thus identify different sites from topography, as higher stacking energy corresponds to greater interlayer repulsion, and increased topographic height.

To further confirm that our device maintains rhombohedral stacking order, we study the spectroscopic signatures of the rhombohedral phase. The major difference between Bernal (ABA) and rhombohedral (ABC) band structure (see Extended Data Fig.~\ref{fig:extended1}) is the energy position of the remote bands; for ABC the remote bands are separated by $2\gamma_1 \;(\approx 0.8~$eV), whereas in ABA they are separated by $2\sqrt{2}\gamma_1 \;(\approx 1.1~$eV) (see SI sec.~S1). Here, $\gamma_1$ is the interlayer hopping between graphene layers.  Fig.~\ref{fig:fig1}f shows the experimentally measured d$I$/d$V$ spectra on a large energy scale at site C$_\mathrm{B}$ (marked by red dot in Fig.~\ref{fig:fig1}e) at gate voltage $V_\text{g}$ = $-32$~V which is the charge neutrality point (CNP). We identify the CNP from the $V_\text{g}$-dependent d$I$/d$V$ spectra that are discussed below. The experimentally observed remote band positions (marked by the two vertical arrows) agree well with the calculated ABC graphene spectra (see Extended Data Fig.~\ref{fig:extended1}). If we zoom in at low energies where the effect of the {\moire} potential is dominant and take d$I$/d$V$ spectra with better energy resolution (inset), we observe two prominent peaks, identified as valence flat band (VFB) and conduction flat band (CFB), intrinsic to the bare rtG. Additionally, there are two smaller peaks identified as {\moire} valence remote band (mVRB) and {\moire} conduction remote band (mCRB), which appear due to band folding from the {\moire} superlattice. The peak widths (using Gaussian fitting) of the VFB and CFB are $4.4$~mV and $7.5$~mV, respectively, and are separated by a gap $\Delta$ of around $19.4$~mV at CNP. We next discuss how these peak features due to VFB, CFB, mVRB, and mCRB evolve with doping and displacement field. 

\begin{figure*}
\includegraphics[width=16.8cm]{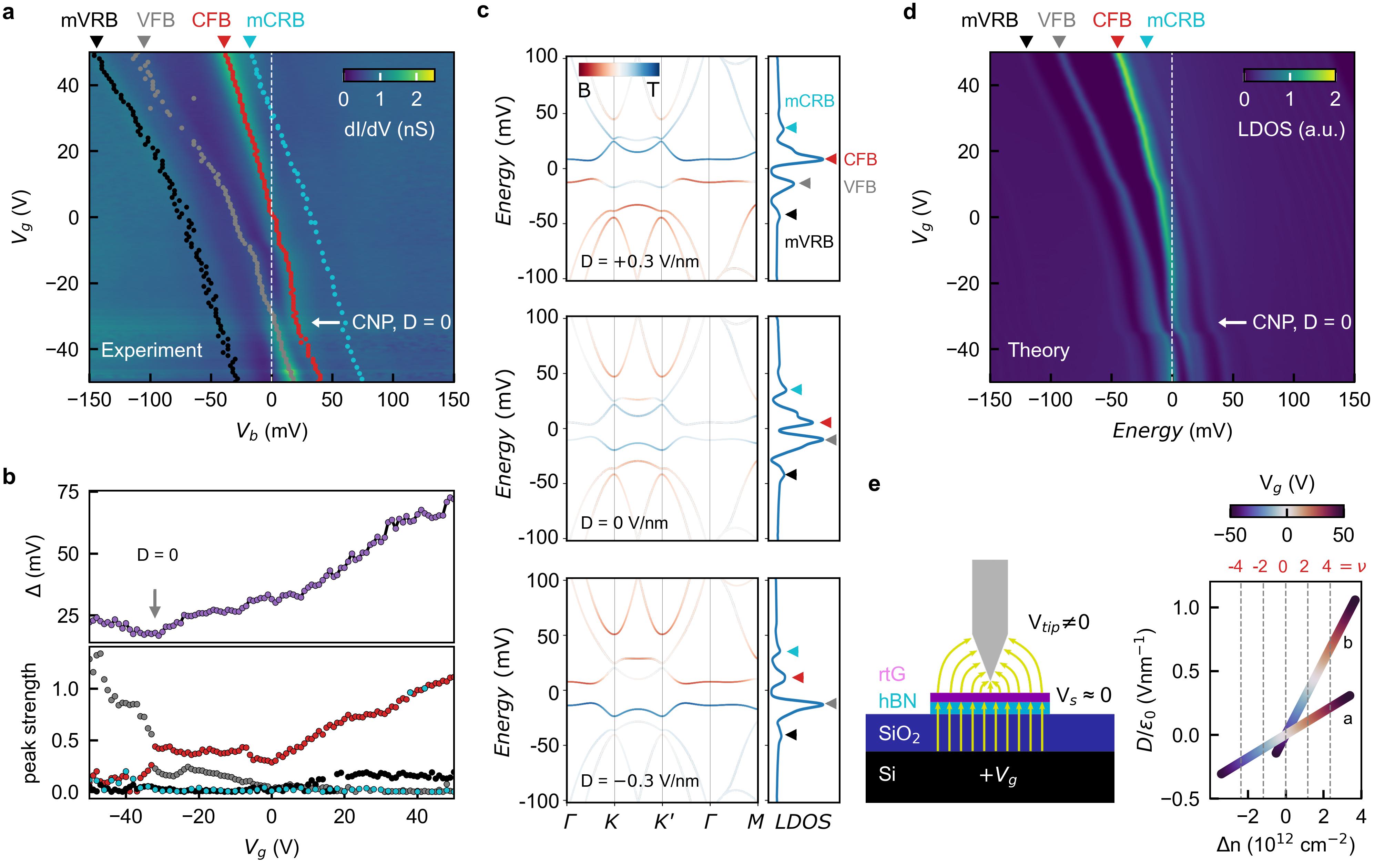}
\caption{ \label{fig:fig2}  \textbf{Evolution and layer polarization of flat bands with gate voltage in trilayer ABC graphene/hBN {\moire}.} (a) Gate voltage ($V_\text{g}$) dependence of d$I$/d$V$ at site C$_\mathrm{B}$ from $-50$~V to $+50$~V. Four main features in the spectra CFB, VFB, mCRB, and mVRB are marked by red, gray, cyan, and black arrowheads, respectively. Charge neutrality point (CNP) and zero displacement field point ($D = 0$) are identified ($V_\text{g} = -32$ V) with the horizontal arrow. The vertical white dashed line marks the Fermi level. (b) Top panel shows evolution of the energy gap ($\Delta$) between VFB (gray curve in (a)) and CFB (red curve in (a)) with $V_\text{g}$. We marked the $D = 0$ point where $\Delta$ is minimum. Bottom panel shows the evolution of the strength of the CFB, VFB, mCRB, and mVRB peaks with $V_\text{g}$.  (c) Layer polarized band structure of the trilayer ABC graphene/hBN {\moire} at finite electric field values $+0.3$ (top), $0$ (middle), $-0.3$ (bottom) V/nm. Right panels show the corresponding LDOS projected on the top layer. (d) Calculated LDOS at site C$_\mathrm{B}$ projected on the top layer. Calculated spectra share many characteristics with the experimental spectra in (a). (e) Left: schematic for tip-induced doping and displacement field during STM/STS measurements. Right: Accessible parameter space using a single gate sample geometry. Line a traces parameter space expected from a ideal parallel plate capacitor geometry. Line b traces the parameter space used to reproduce (d) that fits the experiment in (a).}
\end{figure*}

Fig.~\ref{fig:fig2}a shows an intensity color map of the low energy d$I$/d$V$ spectra at the C$_\mathrm{B}$ site as the back gate voltage, $V_\text{g}$, is swept from $-50$~V to $+50$~V. The range of $V_\text{g}$ is restricted to prevent gate leakage. With $V_\text{g}$, we track the evolution of the four distinct peak features seen in the inset of Fig.~\ref{fig:fig1}f, marked by red (CFB), gray (VFB), black (mVRB) and cyan (mCRB) arrowheads. The white vertical dashed line indicates the Fermi level. We extract the energy spacing $\Delta$ between the VFB (gray trace) and CFB (red trace) and plot it as a function of $V_\text{g}$ in the top panel of Fig.~\ref{fig:fig2}b. The gap $\Delta$ is tuned by the applied displacement field in the system. The condition $D = 0$ occurs at $V_\text{g} = -32$~V where $\Delta$ is minimum. $V_\text{g} = -32$~V is also identified as the CNP based on the observation of the sudden change in slope of the dispersion of flat band peaks with $V_\text{g}$ as the Fermi level goes through the gap and encounters the next flat band. In the bottom panel of Fig.~\ref{fig:fig2}b, we plot the peak strengths of the four peaks. We note that at negative $V_\text{g}$, the VFB intensity (gray) reaches its maximum, while at positive $V_\text{g}$, the CFB intensity (red) reaches its maximum, with both intensities being equal at around $-32$~V. The peak intensities of mVRB and mCRB, on the other hand, remain almost invariant under changing $V_\text{g}$ because they lack strong bottom- or top-layer character. Similar gate dependent spectra taken at C$_\mathrm{BN}$ site of the {\moire} superlattice are shown in the SI sec.~S2. Qualitatively, the gate dispersion is similar for both sites; however, there are subtle differences in their spectral features which will be discussed in detail below.

To understand the origin of the peaks in d$I$/d$V$ and their dispersion with $V_\text{g}$, we compute the low-energy band structure of rtG using a continuum model with the {\moire} potential \cite{Jung2014, jung2015origin} due to alignment with hBN. Fig.~\ref{fig:fig2}c presents the band structure at three representative displacement fields, $D = 0.3~$ (top panel), $0$ (middle panel), and $-0.3$ (bottom panel) V/nm. All the bands have a mixed layer character with contributions from both top (color red) and bottom (color blue) graphene layer with varying degrees of strength. As STM measurements are predominantly sensitive to the top graphene layer, we show the calculated LDOS projected on the top layer in the panels on the right. The two intrinsic flat bands that emerge from rtG subject to a displacement field give rise to two prominent peaks in LDOS (red and gray arrows marking CFB and VFB peaks, respectively), along with two weaker peaks that come from higher energy {\moire} bands in the folded band structure (black and cyan arrows marking {\moire} valence remote band mVRB and {\moire} conduction remote band mCRB, respectively). We identify various peaks in our experimental d$I$/d$V$ spectra (marked by arrows in Fig.~\ref{fig:fig2}a) based on these band structures, and their dispersion with $V_\text{g}$ can be understood in terms of band gaps induced by the {\moire} potential. In our calculations, we assumed the {\moire} strength to be $14.88$~mV \cite{Jung2014, jung2015origin}. Further details on how the band structure changes with the strength of the {\moire} potential and displacement field are provided in the SI sec.~S3.

Comparison with the continuum model enables us to understand several qualitative aspects of the gate dependent data. As we change $V_\text{g}$, the chemical potential of the system shifts due to carrier doping in graphene, shifting the overall energy positions of the flat band peaks. Simultaneously, $D$ changes the energy gap ($\Delta$) between VFB and CFB with a minimum $\Delta$ at $D = 0$ as seen in the top inset of Fig.~\ref{fig:fig2}b, as observed previously  \cite{lui_observation_2011}. 
When the displacement field changes from $+D$ ($+V_\text{g}$) to $-D$ ($-V_\text{g}$), the layer polarization of the band structure flips. The density of states in the top layer is dominated by the contribution from CFB at $+D$ and VFB at $-D$ (Fig.~\ref{fig:fig2}c). This $D$ induced layer polarization explains the d$I$/d$V$ intensity switching between VFB and CFB with $V_\text{g}$, and the nearly identical intensity of CFB and VFB at $D = 0$ (Fig.~\ref{fig:fig2}b bottom panel). The peak heights of CFB and VFB are thus a sensitive probe for layer polarization and displacement fields.

Based on the band structures presented in Fig.~\ref{fig:fig2}c, we calculate LDOS projected onto the top graphene layer at site C$_\mathrm{B}$ as a function of $V_\text{g}$, as shown in Fig.~\ref{fig:fig2}d. To match the calculated LDOS map shown in Fig.~\ref{fig:fig2}d with that of experiment, we assume that doping ($n$) and displacement field ($D$) in trilayer graphene vary linearly with $V_\mathrm{g}$ tracing the line \textbf{b} in the $\Delta n-D$ parameter space, shown in right panel in Fig.~\ref{fig:fig2}e (see SI sec.~S4 for details). In comparison, line \textbf{a} traces the expected parameter space for an ideal parallel plate capacitor geometry of the graphene device with back gate and tip. We attribute an overall offset in $D$ and CNP ($V_\mathrm{g}$ = $-32$~V) from $V_\mathrm{g} = 0$ to the work function difference between the tip and graphene \cite{choi_electronic_2019, xie_spectroscopic_2019}. The deviation from the parallel-plate formula arises due to the shape of the tip, which acts as an electric field concentrator. Notably, the calculated LDOS accurately captures the energy positions of the maxima of the LDOS in the flat and remote bands, suggesting that the single particle band structure is a reasonable starting point to understand the features observed in the parameter space explored in our sample. From the observed {\moire} periodicity $\lambda$ ($\approx$ 14~nm), we calculate the moire unit-cell area $A$ ($= (\sqrt{3}/2) \lambda^2$) and {\moire} filling factor $\nu = 4n/n_s$ with $n_s$ representing the number of carriers to fill one {\moire} band and the factor 4 is due to spin and valley degeneracy. The values of $n$ corresponding to different $\nu$ are indicated by vertical dotted lines in Fig.~\ref{fig:fig2}e. We note that tip-induced offsets and gate leakage limit our measurements largely to the electron-doped side of the phase space, corresponding to positive $D$ values.

\begin{figure*}
\includegraphics[width=12.7cm]{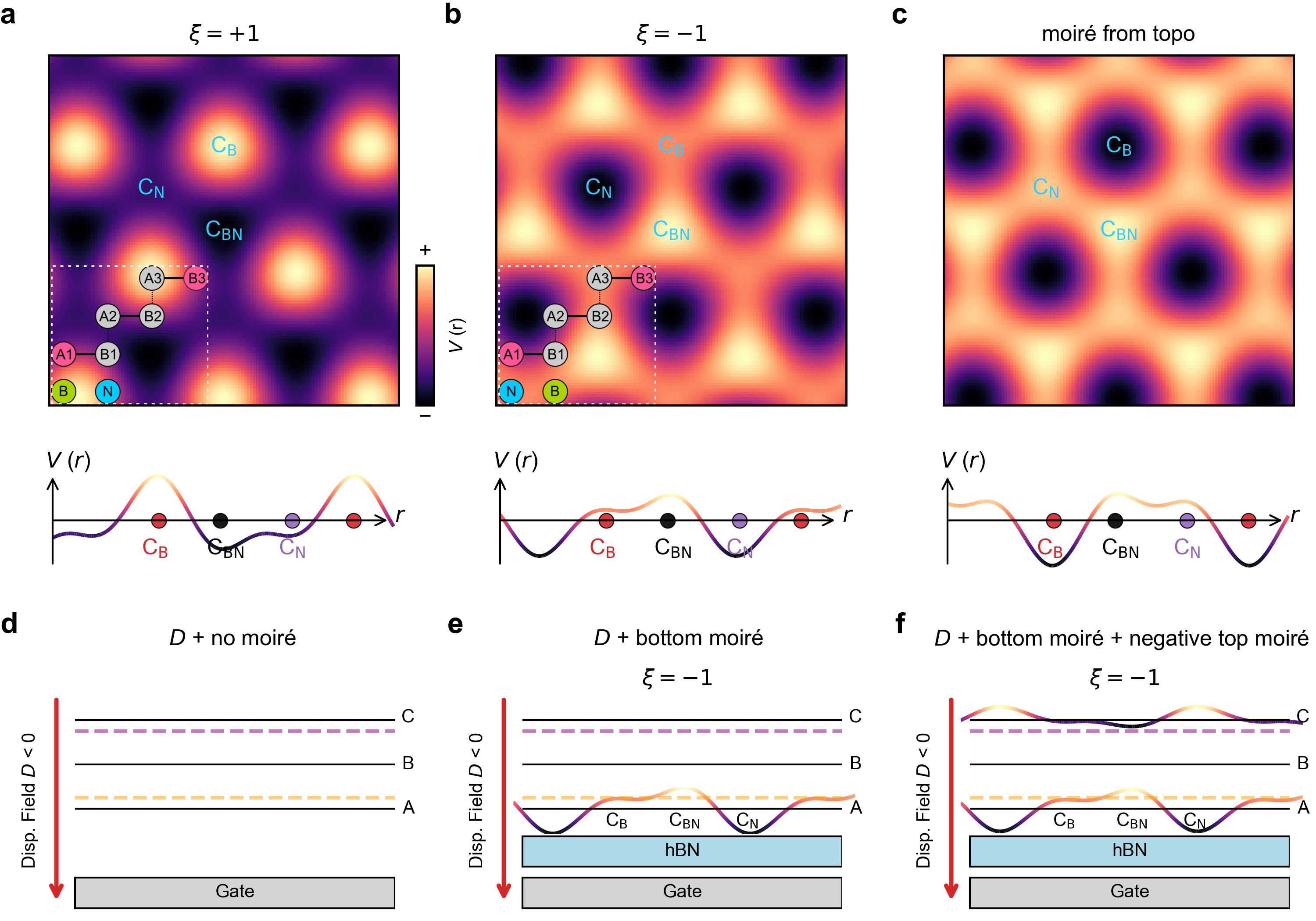}
\caption{ \label{fig:fig3}  \textbf{Real space effect of {\moire} potential as local displacement field.} (a)-(c) {\Moire} potential profiles corresponding to hBN alignments $\xi = \pm 1$, and a {\moire} potential extracted from topography. The high symmetry {\moire} stacking sites C$_\mathrm{BN}$, C$_\mathrm{B}$, and C$_\mathrm{N}$ are labeled. The atomistic hBN alignment with ABC graphene for $\xi = \pm 1$ are shown in the insets to (a), and (b). Low-energy sublattice of bottom layer graphene, A1 sits on boron for $\xi = 1$ and on nitrogen for $\xi = -1$. Lower panels show linecuts of the {\moire} potential $V(r)$ along the high symmetry sites. The orange color denotes positive value of the {\moire} potential and purple color denote negative value. (d) Under applied $-D$ and in absence of {\moire}, constant potential landscape in top and bottom graphene. The constant positive potential at bottom layer is denoted by orange line where as the constant negative potential on top layer is denoted by purple line. (e) In presence of {\moire} on bottom graphene layer the potential landscape modulates with {\moire}. At some sites the potential due to applied $-D$ adds with the {\moire} and at other sites they cancel partially. (f) Potential profile on botttom and top graphene layers when an additional negative {\moire} is imposed on the top graphene layer.}
\end{figure*}

We next consider the real space variation of local density of states and the nature of the {\moire} potential. The physical origin of the {\moire} potential lies in the spatial variation of the electrostatic interaction between carbon atoms in the bottom layer of rtG and boron/nitrogen atoms in the topmost hBN layer, depending on the local stacking arrangements at C$_\mathrm{BN}$, C$_\mathrm{B}$, and C$_\mathrm{N}$. The resulting effect is typically modeled using a single-shell approximation characterized by two parameters: $C$, which controls the overall strength of the potential, and $\phi$, which governs the relative amplitudes at the high-symmetry points (C$_\mathrm{BN}$, C$_\mathrm{B}$, and C$_\mathrm{N}$) \cite{Jung2014, jung2015origin} (see SI sec.~S5). In ABC-stacked rhombohedral trilayer graphene, the low-energy band structure is primarily derived from the A1 sublattice in the bottom layer and the B3 sublattice in the top layer (inset to Fig.~\ref{fig:fig3}a), while the remaining orbitals carry significant spectral weight only at energies $\sim 400$~meV above the charge neutrality point. For a given {\moire} stacking site, such as C$_\mathrm{BN}$ (insets to Fig.~\ref{fig:fig3}a,b), two distinct alignment orientations of rtG with hBN are possible \cite{Chittari2019, Gonzalez2021a, Park2023}, corresponding to a new parameter $\xi = \pm 1$ depending on whether the A1 sublattice sits on boron ($\xi = +1$) or nitrogen ($\xi = -1$). These two orientations are related by a global $180^\circ$ rotation of the hBN lattice and thus can not be accessed in a single device. More recently, this sublattice-specific alignment has been shown to influence correlated phases in pentalayer rhombohedral graphene \cite{Uzan2025_hBNalignment}.

In Fig.~\ref{fig:fig3}a-c, we present three representative {\moire} potentials $V(r)$: $\xi = +1$ ($C = -14.88$~mV, $\phi = 50.19^\circ$), $\xi = -1$ ($C = 12.09$~mV, $\phi = -46.64^\circ$), corresponding to the two possible microscopic hBN alignments, and a third case ($C = 15$~mV, $\phi = 54.64^\circ$) where the parameters are chosen to match the experimentally measured topographic height at the high-symmetry stacking sites (see SI sec.~S6). As shown in Fig.~\ref{fig:fig3}a–c, each stacking site experiences a distinct local potential for a given {\moire} potential profile. 
To understand the role of the {\moire} potential, first consider the case without it.
The displacement field induces a uniform interlayer potential difference between the bottom (A) and top (C) graphene layers, schematically shown in Fig.~\ref{fig:fig3}d. This induces a spatially uniform layer polarization in which the relative peak amplitudes of the CFB and the VFB depend sensitively on the displacement field (see Fig.~\ref{fig:fig2}). For instance, at $D < 0$ with no {\moire}, the CFB/VFB peak ratio is uniform in space, with VFB having a higher peak intensity. 

\begin{figure*}
\includegraphics[width=16.8cm]{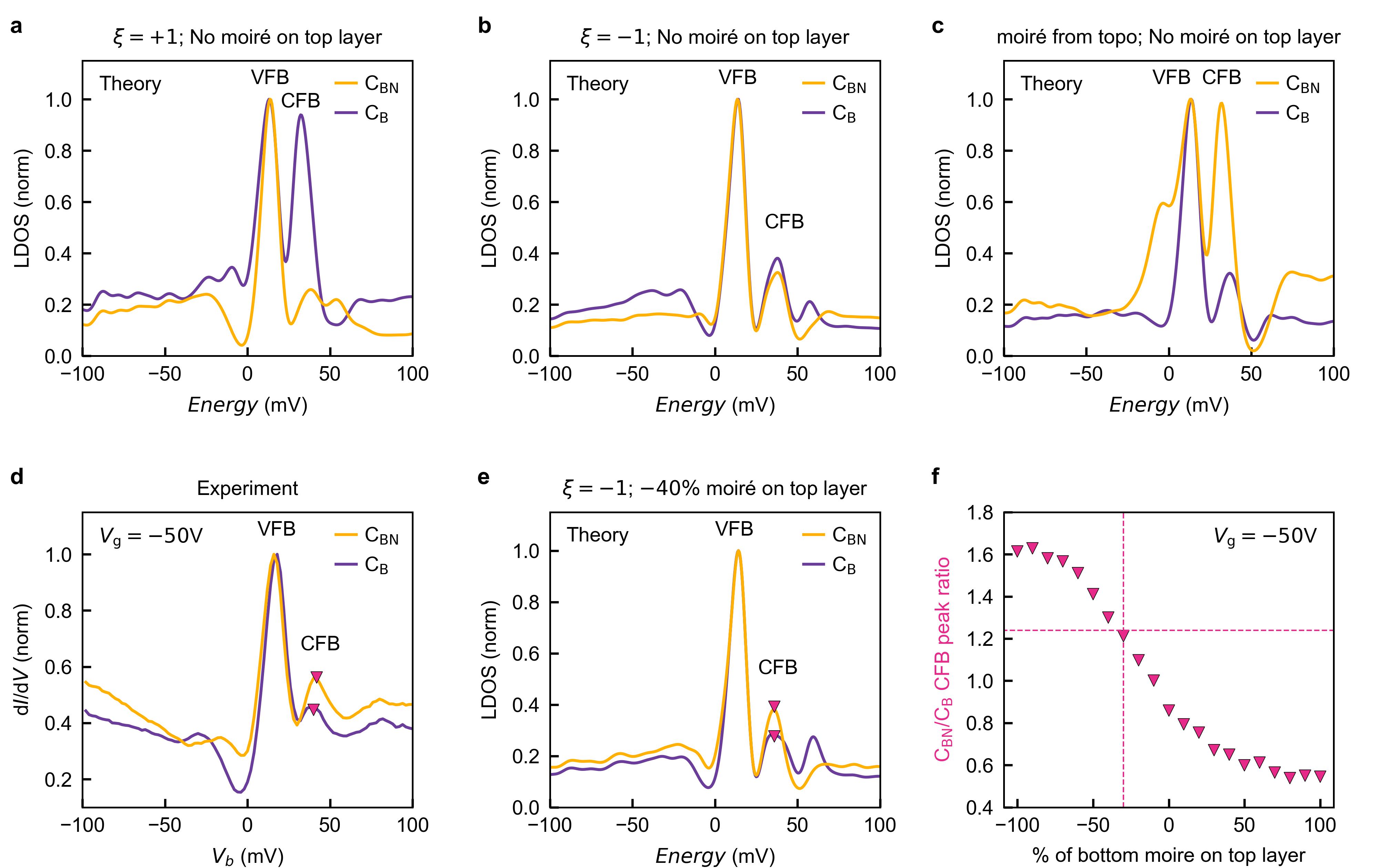}  
\caption{ \label{fig:fig4} \textbf{Theoretical real space LDOS variation for different {\moire} potentials and comparison with experiment}
(a)-(c) Calculated LDOS spectra at C$_\mathrm{BN}$ and C$_\mathrm{B}$ sites at $V_g = -50$ V for three different {\moire} potentials corresponding to hBN alignments $\xi = +1$, $\xi = -1$, and a {\moire} mimicking topography, respectively. The {\moire} potentials act only on the bottom layer of rhombohedral trilayer graphene (rTG). All the spectra are normalized by the VFB intensity. (d) Experimentally measured d$I$/d$V$ spectra at C$_\mathrm{BN}$ and C$_\mathrm{B}$ sites for $V_g = -50$. The CFB peaks are marked by red triangles. The measured flat band intensities better match the calculated LDOS spectra in (b) for $\xi = -1$. (e) Calculated LDOS spectra with $\xi = -1$ {\moire} when a negative ($-40 \%$) $\xi = -1$ {\moire} potential is additionally imposed onto the top graphene layer. The CFB peaks and {\moire} induced dips are again marked by red triangles. (f) Calculated C$_\mathrm{BN}$/C$_\mathrm{B}$ CFB peak ratio as a function of the percentage of the $\xi = -1$ {\moire} potential applied to the top layer. The dashed lines indicate where the theoretical modeling fits to experiment.
}
\end{figure*}

Adding a {\moire} potential at the rtG/hBN interface spatially modulates the layer polarization and consequently the relative CFB/VFB intensities. This is reflected in the continuum model, where the {\moire} potential enters the Hamiltonian as a diagonal on-site term and can be treated as an effective local displacement field. We show schematically the scenario when a $\xi = -1$ {\moire} potential is introduced at the rtG/hBN interface in Fig.~\ref{fig:fig3}e. At sites C$_\mathrm{BN}$ and C$_\mathrm{B}$, the {\moire} potential is positive and partially cancels out the negative potential due to applied $D$. The inverse is true at C$_\mathrm{N}$ where the {\moire} potential is negative and the local potential is reinforced.
The modulation of flat-band intensity arising from local variations in $D$ at different stacking sites thus provides a probe of the {\moire} potential profile. 
As we will see next in Fig.~\ref{fig:fig4}, a {\moire} potential on the bottom-layer alone cannot explain the site-dependent variations in the measured d$I$/d$V$ and necessitates a {\moire} potential on top layer, specifically one with an opposite sign to the bottom layer. The effect of such a potential is schematically shown in Fig.~\ref{fig:fig3}f.

Fig.~\ref{fig:fig4}a–c show the calculated LDOS spectra at C$_\mathrm{BN}$ and C$_\mathrm{B}$ projected onto the top layer for three types of {\moire} potentials at $V_\mathrm{g} = -50$~V (n.b. all spectra are normalized to the VFB peak). Because the CFB is polarized toward the bottom layer, $\xi = +1$ leads to constructive addition at C$_\mathrm{B}$, yielding a stronger CFB peak than at C$_\mathrm{BN}$. Similar trends apply for the other {\moire} potentials.
We now compare the calculated LDOS spectra with the experimental STS (Fig.~\ref{fig:fig4}d) at sites C$_\mathrm{BN}$ and C$_\mathrm{B}$. Experimentally, the CFB peaks have comparable amplitudes, with C$_\mathrm{BN}$ higher by a factor of 1.24. Among the three {\moire} potentials, $\xi = -1$ best reproduces the comparable CFB intensities at the C$_\mathrm{BN}$ and C$_\mathrm{B}$ sites observed in the experiment. The other two potentials largely overestimate the layer polarization asymmetry between the C$_\mathrm{BN}$ and C$_\mathrm{B}$ spectra.

While the $\xi = -1$ {\moire} potential reproduces several features of the experimental spectra, it does not capture the relative CFB peak heights accurately.
Experimentally, we find that the CFB peak is lower at C$_\mathrm{B}$ than at C$_\mathrm{BN}$, indicating a relatively stronger layer polarization for the CFB at C$_\mathrm{B}$, which is opposite to the trend for $\xi = -1$ in Fig.~\ref{fig:fig4}c. 
This discrepancy can be resolved by introducing a {\moire} potential on the top layer.
The effect of such a potential on the C$_\mathrm{BN}$/C$_\mathrm{B}$ CFB peak-height ratio is shown in Fig.~\ref{fig:fig4}f. 
When the top-layer {\moire} potential has the same sign as the bottom layer, it reinforces the existing polarization asymmetry between the C$_\mathrm{BN}$ and C$_\mathrm{B}$ sites. In contrast, a negative top-layer potential opposes this asymmetry by offsetting the bottom-layer contribution in the layer-projected LDOS. Quantitatively, a top layer {\moire} potential of $\sim -35\%$ of the bottom layer value matches the experimental CFB peak ratio (magenta triangle in Fig.~\ref{fig:fig4}e,f) at $V_{g} = -50 V$. Similar result holds for $D>0$ ($V_\mathrm{g} = -20$ V) as shown in SI sec.~S7.
While this value is sensitive to the gate voltage (see SI sec.~S8 for details), we find that
a negative top-layer {\moire} potential of at least $30\%$ is required to reproduce the peak ratios in the spectra. Interestingly, recent Hartree–Fock studies have also suggested such {\moire} potential renormalizations \cite{Kwan2025}.

We have developed a fabrication process capable of producing STM-quality, few-layer graphene devices with large, well-defined regions of rhombohedral stacking. This platform enables detailed characterization of the electronic structure of the ABC trilayer graphene/hBN {\moire} superlattice at liquid helium temperatures. In addition to the intrinsic flat bands of rtG, we observe {\moire}-induced gaps openings and remote bands, revealing a spatially modulated electronic structure. By comparing experimental LDOS with continuum model calculations, we outline the sensitivity of the electronic structure to the relative orientation of hBN, with the $\xi = -1$ alignment yielding better agreement with experiment, particularly when a negative
{\moire} potential is imposed upon the top layer as well. Moreover, we find that the {\moire} potential strength on the top layer varies with gate voltage, a behavior that is difficult to reconcile within a purely single-particle framework.
Looking ahead, this technique paves the way for direct imaging of correlated quantum phases in multilayer rtG/hBN heterostructures at mK temperatures, as well as for exploring the effects of spin–orbit coupling via proximity to transition metal dichalcogenides such as $\mathrm{WSe_2}$.

%-------------------------------------------------------------------

\bibliography{abc}

\clearpage

%----------------------------------------------------------------------

\clearpage

%------------------   Extended Figures   ---------------------

\setcounter{figure}{0}

\renewcommand{\figurename}{\textbf{Extended Data Fig.}}

 \makeatletter
\def\@fnsymbol#1{\ensuremath{\ifcase#1\or \dagger\or *\or \ddagger\or
   \mathsection\or \mathparagraph\or \|\or **\or \dagger\dagger
   \or \ddagger\ddagger \else\@ctrerr\fi}}
    \makeatother

\setcounter{figure}{0}
\renewcommand{\figurename}{\textbf{Extended Data Fig.}}

\begin{figure*}
\includegraphics[width=16.8cm]{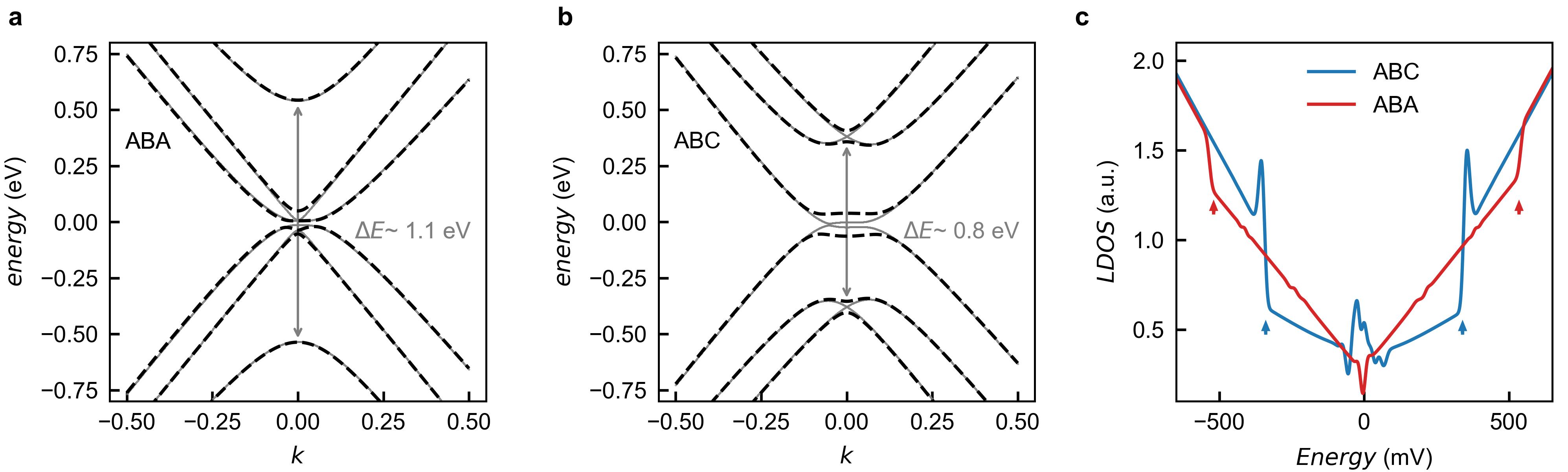}
\caption{\label{fig:extended1} \textbf{Band structure of ABA and ABC trilayer graphene.} (a), (b) Tight-binding band structure of ABA and ABC trilayer graphene, respectively. The solid gray curves are at zero displacement field, the black dashed curves are for a non-zero displacement field that opens up a gap in the system at band crossings. The key difference between ABA and ABC graphene is the energy separation between the higher-energy remote bands. (c) Calculated local density of state (LDOS) of ABA and ABC graphene. Arrows mark the positions of the remote bands. For ABC, the remote bands are separated by $0.8$~eV and for ABA the remote bands are separated by $1.1$~eV.  }

\end{figure*}

%--------------------------------------------------------------------

\clearpage

\begin{acknowledgments}
We thank Bhima L. Chittari for insightful discussions on hBN alignment. This work was supported by Programmable Quantum Materials, an Energy Frontier Research Center funded by the U.S. Department of Energy (DOE), Office of Science, Basic Energy Sciences (BES) under award no. DE-SC0019443. The Flatiron Institute is a division of the Simons Foundation.
\end{acknowledgments}

\section*{Author's contribution}
E.S.~and S.G.~performed STM experiments and analysis. 
E.S.~and N.V.~performed theoretical calculations. J.C.~fabricated the sample and performed Raman characterization. 
B.Y. helped in fabrication. Y.D. performed SNOM imaging.
T.T.~and K.W.~provided the hBN crystals.
A.B. and M.M.D. provided SiN masks for contact deposition.
E.S., S.G., N.V., J.C., and A.N.P. wrote the paper, with input from all authors. 
A.N.P., D.N.B, R.Q. and C.D. supervised the project. 
All authors discussed the results.

\section*{Methods}

\subsection{Device fabrication}

Polypropylene carbonate films were spin coated onto SiO2/Si substrates using a spin speed of 3450 rpm and subsequently baked at $120^\circ$~C for five minutes. Bulk graphite crystals were cleaved on tape and deposited on the PPC-covered substrates using standard mechanical exfoliation practices. The entire process was performed with the substrates heated to $~50^\circ$~C, around the glass transition of PPC where adhesion is maximized. Thin flakes were identified using optical contrast and confirmed by Raman spectroscopy, which further identified their stacking order. After identifying suitable flakes, the PPC film was removed using sticky tape with a hole and placed onto a flat PDMS cylinder. Then, the flakes were deposited onto hBN using a dry transfer technique. After transfer, large areas of the flake were cleaned using contact mode atomic force microscopy scans at a low setpoint of $\approx$ 0.08 V, and electrical contacts consisting of Cr/Au (10/70 nm) were made using a SiN shadow mask. Finally, the stacking order and homogeneity were confirmed using Raman and SNOM, respectively.

\subsection{Measurements}

\subsubsection{Scanning tunneling microscopy/spectroscopy (STM/STS)}

STM/STS measurements were performed using a home-built $7$~K system at ultrahigh vacuum. Prior to scanning, the sample was annealed at  $\approx$ $200^\circ$~C in ultrahigh vacuum to remove surface contaminants. All measurements were taken using an electrochemically etched tungsten tip whose atomic sharpness and electronic behavior were calibrated using single crystal Au (111). Multiple independently prepared tips were used to confirm the consistency of results. 

\subsubsection{Raman spectroscopy}

Raman measurements were performed using a commercial Renishaw inVia confocal Raman microscope. A laser wavelength of 532 nm and excitation power of  $\approx$ 400 $\mu$W were used to minimize potential heating and reversion of the rhombohedral domains. 

\section*{Data availability}

The additional data supporting the findings of this study are available from the corresponding authors upon reasonable request.

%-------------------------------------------------------------------

\clearpage

\end{document}